\documentclass[12pt]{iopart}

\usepackage[english]{babel}
\usepackage[utf8]{inputenc}
\usepackage[T1]{fontenc}

\expandafter\let\csname equation*\endcsname\relax
\expandafter\let\csname endequation*\endcsname\relax
\usepackage{amsmath}
\usepackage{xcolor}
\usepackage[colorlinks,allcolors=violet]{hyperref}
\usepackage{cleveref}
\usepackage{graphicx}
\usepackage{url}
\usepackage{booktabs}
\usepackage{glossaries}
\usepackage{xspace}
\usepackage{tabularx}
\usepackage[
    retain-unity-mantissa=false,
    separate-uncertainty=true
]{siunitx}
\usepackage{subcaption}

\begin{document}

\title[LISAmax: Improving the Low-Frequency Gravitational-Wave Sensitivity]{LISAmax: Improving the Low-Frequency Gravitational-Wave Sensitivity by Two Orders of Magnitude}

\author{W.~Martens\textsuperscript{1}, M.~Khan\textsuperscript{1}, and J.-B.~Bayle\textsuperscript{2}}

\address{\textsuperscript{1}Mission Analysis Section, European Space Agency, Darmstadt, Germany}
\address{\textsuperscript{2}University of Glasgow, Glasgow G12 8QQ, United Kingdom}

\ead{waldemar.martens@esa.int}

\begin{abstract}
Within its Voyage 2050 planning cycle, the European Space Agency (ESA) is considering long-term large class science mission themes. Gravitational-wave astronomy is among the topics under study. Building on previous work by other authors~\cite{folkner2011non,ni2010gravitational,sesana2021unveiling}, this paper studies a gravitational-wave interferometer concept, dubbed ``LISAmax'', consisting of three spacecraft, each located close to one of the Sun-Earth libration points L3, L4 and L5, forming a triangular constellation with an arm length of 259 million kilometers (to be compared to LISA's 2.5 million kilometer arms). We argue that this is the largest triangular formation that can be reached from Earth without a major leap in mission complexity and cost (hence the name). The sensitivity curve of such a detector is at least two orders of magnitude lower in amplitude than that of LISA, at frequencies below \SI{1}{\milli\hertz}. This makes the observatory sensitive to gravitational waves in the \si{\micro\hertz} range and opens a new window for gravitational-wave astronomy, not covered by any other planned detector concept. We analyze in detail the constellation stability for a 10-year mission in the full numerical model including insertion, dispersion, and self-gravity-induced accelerations. We compute the orbit transfers using a European launcher and chemical propulsion. Different orbit options, such as precessing, inclined orbits, the use of flybys for the transfer, and the launch strategy, are discussed. The payload design parameters are assessed, and the expected sensitivity curve is compared with a number of potential gravitational-wave sources. No show stoppers are identified at this point of the analysis.
\end{abstract}


\submitto{\CQG}

\section{Introduction}

In June 2021, the European Space Agency (ESA) has chosen the future science mission themes for the time frame 2035-2050. These are summarized under the so-called Voyage 2050 program~\cite{voyage2050-intro}. Gravitational waves are among the chosen science themes covered by the L6 slot ``New Physical Probes of the Early Universe''~\cite{voyage2050-result}. This has triggered some interest in studying successor space missions to the Laser Interferometer Space Antenna (LISA) currently planned to be launched in 2035~\cite{DanzmannLisaProposal,GravitationalUniverse}. Four White Papers related to gravitational-wave astronomy have been submitted to the Voyage 2050 call~\cite{sesana2021unveiling,Sedda_2021,Baibhav_2021,baker2021high} targeting different frequency ranges. Moreover, there are several space-based gravitational-wave observatory proposals and projects at different stages of development~\cite{folkner2011non,ni2010gravitational,Kawamura_2008,Luo_2016,phinney2004big,crowder2005beyond,taiji,lagrange,cornish2002journey,husa2000,NI2013525, ni2013gravitational}. Each of them aims at either a better sensitivity or a different frequency band compared to LISA. All of these use one or more triangular spacecraft constellations located at different regions of space. 

The current work studies a potential LISA successor targeting the \si{\micro\hertz} band. The study is driven by the question: how large can a post-LISA space-based gravitational-wave observatory be, while still remaining in the budget of an L-class mission (i.e., around one billion Euros at today's costing)? The constellation size, or arm length, mainly determines the sensitive frequency band. Larger arm lengths imply lower frequencies, while also improving the sensitivity in general. Such an observatory would, for instance, allow the observation of super-massive black hole inspirals long before they enter the LISA band. For a more in-depth survey of the science case of a \si{\micro\hertz} observatory, refer to~\cite{sesana2021unveiling}.

To target a low frequency range, one can start by placing three spacecraft at the triangular liberation points L3, L4, and L5 of the Sun-Earth system as proposed in ~\cite{NI2013525,ni2013gravitational,MEN2010434}. These are equilibrium points located at \SI{1}{\astronomicalunit} from the Sun, which span an equilateral triangle. The downside of such a constellation is, however, that L3 is located behind the Sun as seen from Earth. This would preclude direct communications from Earth and require one of the other satellites to act as a relay, increasing the mission complexity. A more robust solution to this problem is to rotate the full constellation by \SI{5}{\degree} to \SI{10}{\degree}, allowing a direct link to Earth for all three satellites. It turns out that such a rotation does not degrade the constellation stability in a significant way. Such an offset triangular constellation was first used in reference~\cite[fig. 2]{folkner2011non}. However, the aim in that proposal was not to target lower frequencies but to recover part of the LISA performance by a simpler, non-drag-free satellite concept. An \si{\astronomicalunit}-sized drag-free detector consisting of two constellations that aims at the \si{\micro\hertz} range was later proposed in~\cite{sesana2021unveiling}. This reference does not provide many details on the orbital mechanics aspects but focuses on the science case of a \si{\micro\hertz} detector. We note, however, that the out-of-plane constellation of that proposal is not reachable with today's propulsion systems due to the high $\Delta v$ demands. The present paper is an attempt to condense the proposed concepts into an operationally feasible and realistic mission scenario.

The most constraining aspect of the studied detector, dubbed ``LISAmax'', is that the constellation is bound to the ecliptic plane. Even though the three satellites rotate around the Sun once per year, this limits the antenna pattern and thus possibly the directional sensitivity of the observatory. Placing the constellation out of the ecliptic plane would require prohibitive amounts of $\Delta v$ during the transfer. Higher inclinations can be reached using planetary flybys. However, reaching such inclinations and then accurately maintaining the triangular configuration would imply a major leap in mission complexity and cost.

The rest of the paper is organized as follows: the design of the science orbit in the full dynamical model and the analysis of its stability under dispersions are presented in \cref{sec:science-orbit,sec:monte-carlo}. \Cref{sec:sensitivity} follows with a discussion of the achievable sensitivity curve. In \cref{sec:transfer}, the interplanetary transfer to the science orbit is analyzed. Various mission options of using inclined, precessing orbits or planetary flybys are discussed in \cref{sec:options}. \Cref{sec:conclusions} concludes the paper.

\section{LISAmax science orbit}
\label{sec:science-orbit}

For the design of the LISAmax science orbit, we assume similar requirements as for LISA~\cite{martens2021trajectory} extending into LISAmax frequency range.

\begin{enumerate}
    \item The total duration of the science phase is 10 years including extension, i.e. the orbits need to be stable for this duration.
    \item No station keeping maneuvers are assumed during the science phase. Such maneuvers would interrupt the science operations and are followed by a re-acquisition of the constellation, which is not desirable.
    \item The corner angles of the constellation during 10 years shall not deviate from the nominal value of \SI{60}{\degree} by more than \SI{+-1.0}{\degree}. This constraint comes mainly from the limitation of the optical assembly tracking mechanism (OATM), which ensures that the laser beams point to the correct direction at all times.
    \item The relative velocity (arm length rate of change) shall not exceed \SI{10}{\meter\per\second} during the 10-year science phase duration. This is due to the bandwidth limitation of the phasemeter that detects the beat notes in the interferometric laser signal~\cite{Phasemeter}.
\end{enumerate}

\begin{figure}[h!]
    \centering
    \includegraphics[width=\linewidth]{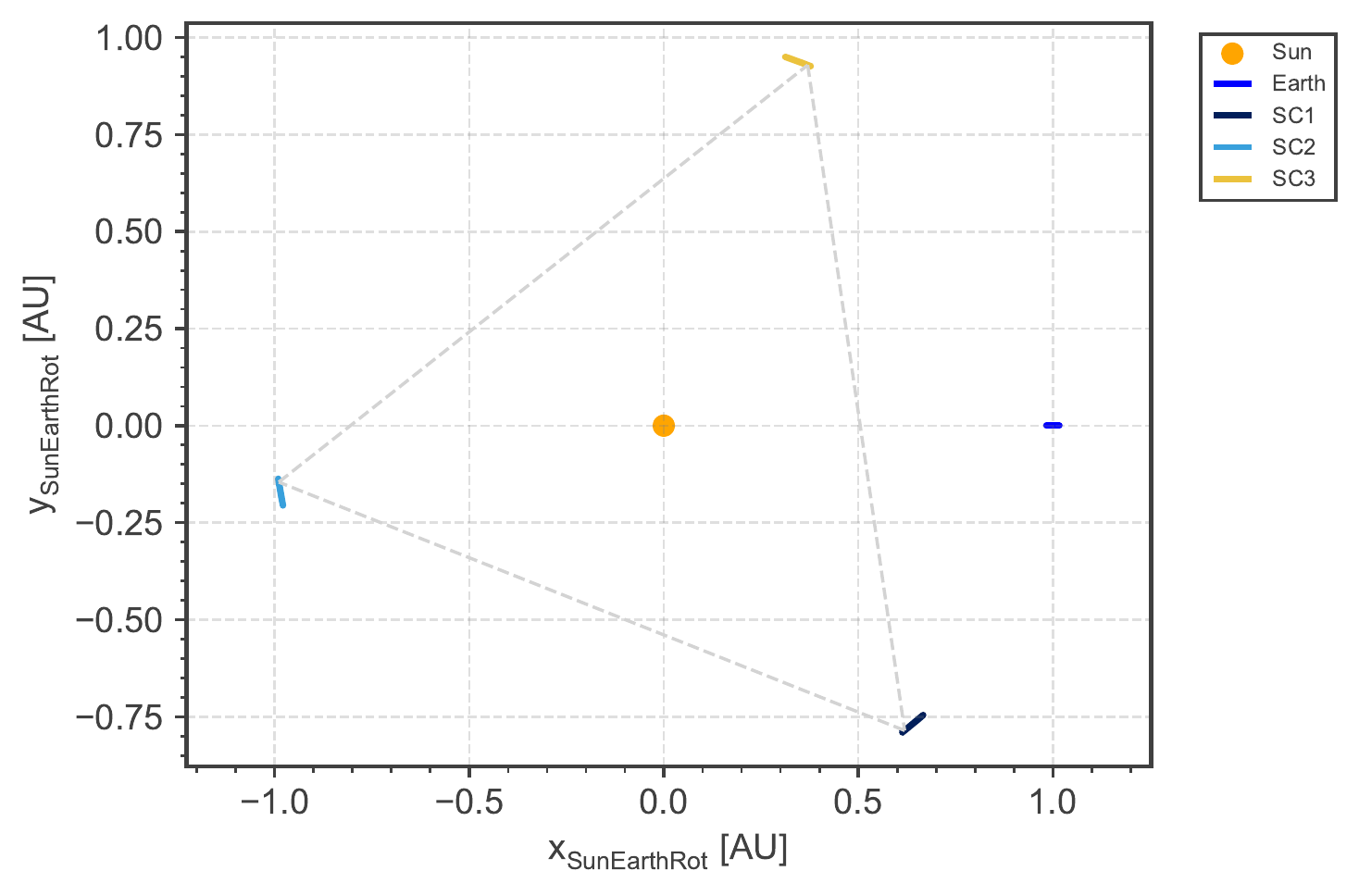}
    \caption{The optimized science orbit of LISAmax over a 10-year duration in the Sun-Earth rotating frame. The elongated shape of the Earth and spacecraft trajectories in the plot indicate the variation of their locations over the science phase duration.}
    \label{fig:science_orbit}
\end{figure}

\begin{figure}[h!]
    \centering
    \begin{subfigure}{\linewidth}
        \includegraphics[width=\linewidth]{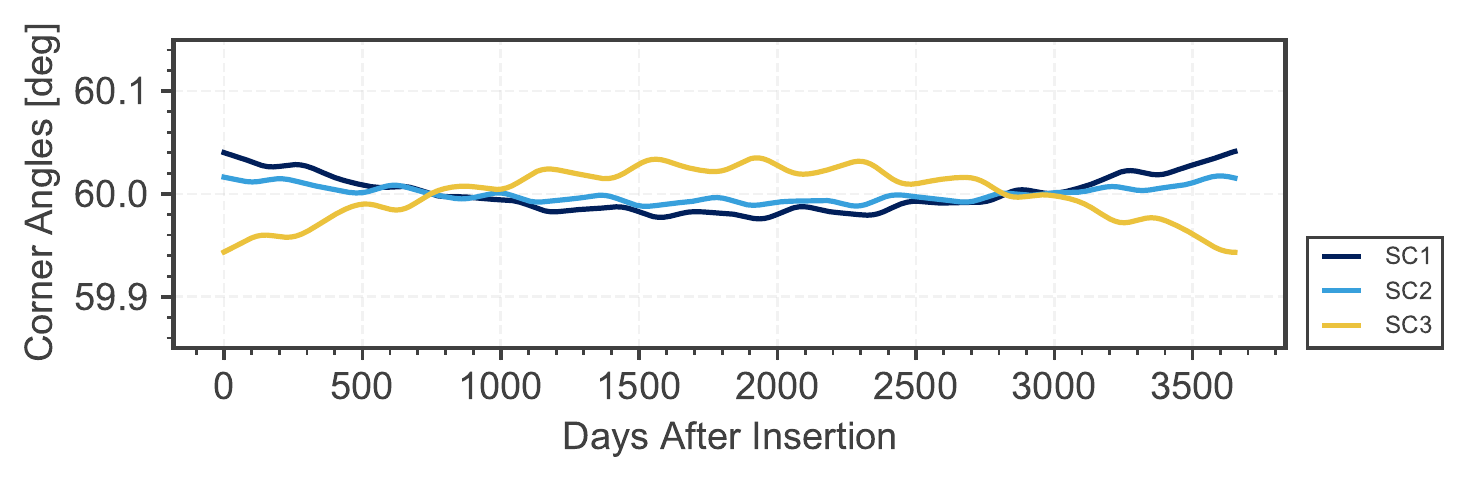}
        \caption{}  
    \end{subfigure}
    \begin{subfigure}{\linewidth}
        \includegraphics[width=\linewidth]{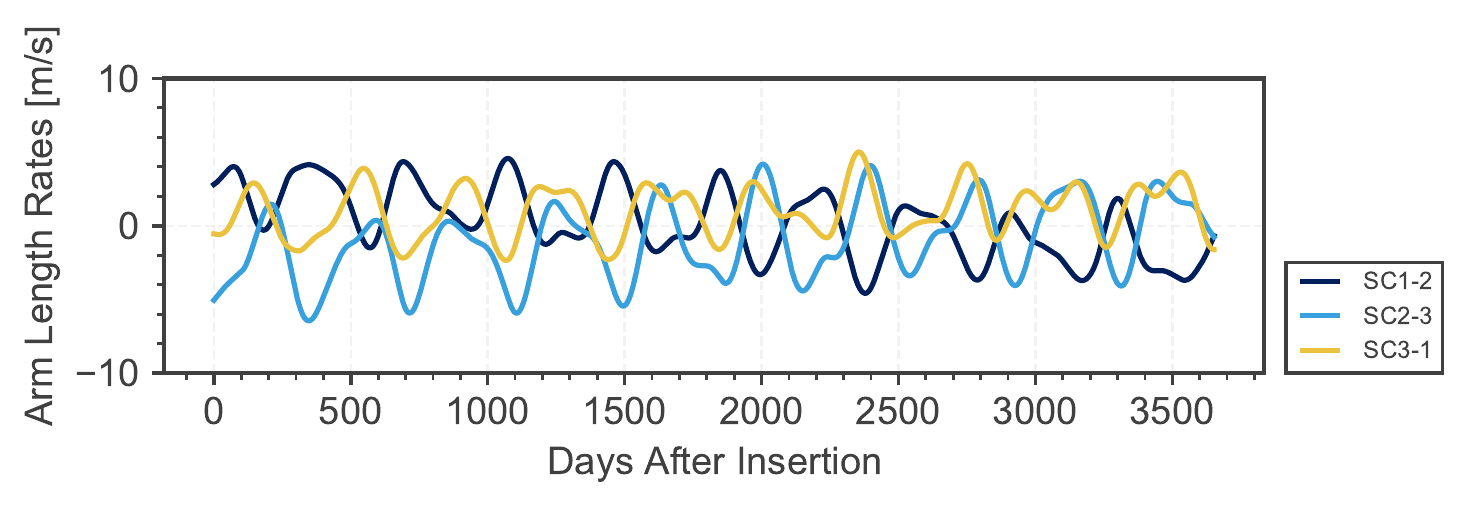}
        \caption{}  
    \end{subfigure}
    \caption{Corner angles (a) and arm length rates (b) evolution over the 10-year science phase duration.}
    \label{fig:orbit_geometry}
\end{figure}

It is convenient to represent the orbits of LISAmax in the Sun-Earth rotating frame, with the $x$-axis along the Sun-Earth line; the $z$-axis along the Earth orbit angular momentum vector; and the $y$-axis completing the right-handed system. The fully optimized science orbit in this frame is depicted in~\cref{fig:science_orbit}. The gravitational accelerations of all relevant celestial bodies (Sun, Earth, Moon, Venus, Mars, Jupiter, Saturn) have been taken into account in the numerical propagation using a Runge-Kutta integrator. Solar radiation pressure is not affecting the spacecraft trajectories because its effect is assumed to be compensated by the Drag Free and Attitude Control System (DFACS), similarly to LISA~\cite{gath2007drag,klotz1997drag,vidano2020lisa}. The problem was set up as a constraint satisfaction problem using GODOT~\cite{godot} for the trajectory propagation, PyGMO~\cite{Biscani2020} and the NLP solver WORHP~\cite{buskens2013esa} for optimization. As an initial guess, Keplerian orbits in a triangular formation, offset by \SI{10}{\degree} from the Sun-Earth triangular libration points, have been used. Constraints have been imposed on the maximum deviation of all three corner angles, arm lengths, and arm length rates. The maximum values for these constraints have been manually adjusted to be as small as possible while still allowing for convergence of the NLP solver. The final values of the maximum deviations are \SI{60+-0.06}{\degree} for the corner angles and \SI{+-5.0}{\meter\per\second} for the arm length rates. The full evolution of both quantities is shown in ~\cref{fig:orbit_geometry}. These variations are significantly smaller that what is achievable for LISA~\cite{martens2021trajectory} (\SI{60+-1.0}{\degree} for the corner angles and \SI{+-10.0}{\meter\per\second} for the arm length rates). This is due to the larger distance from perturbing sources, mainly the Earth.

\section{Science orbit insertion accuracy and self-gravity-induced accelerations}\label{sec:monte-carlo}

\begin{figure}[h!]
    \centering
    \includegraphics[width=7cm]{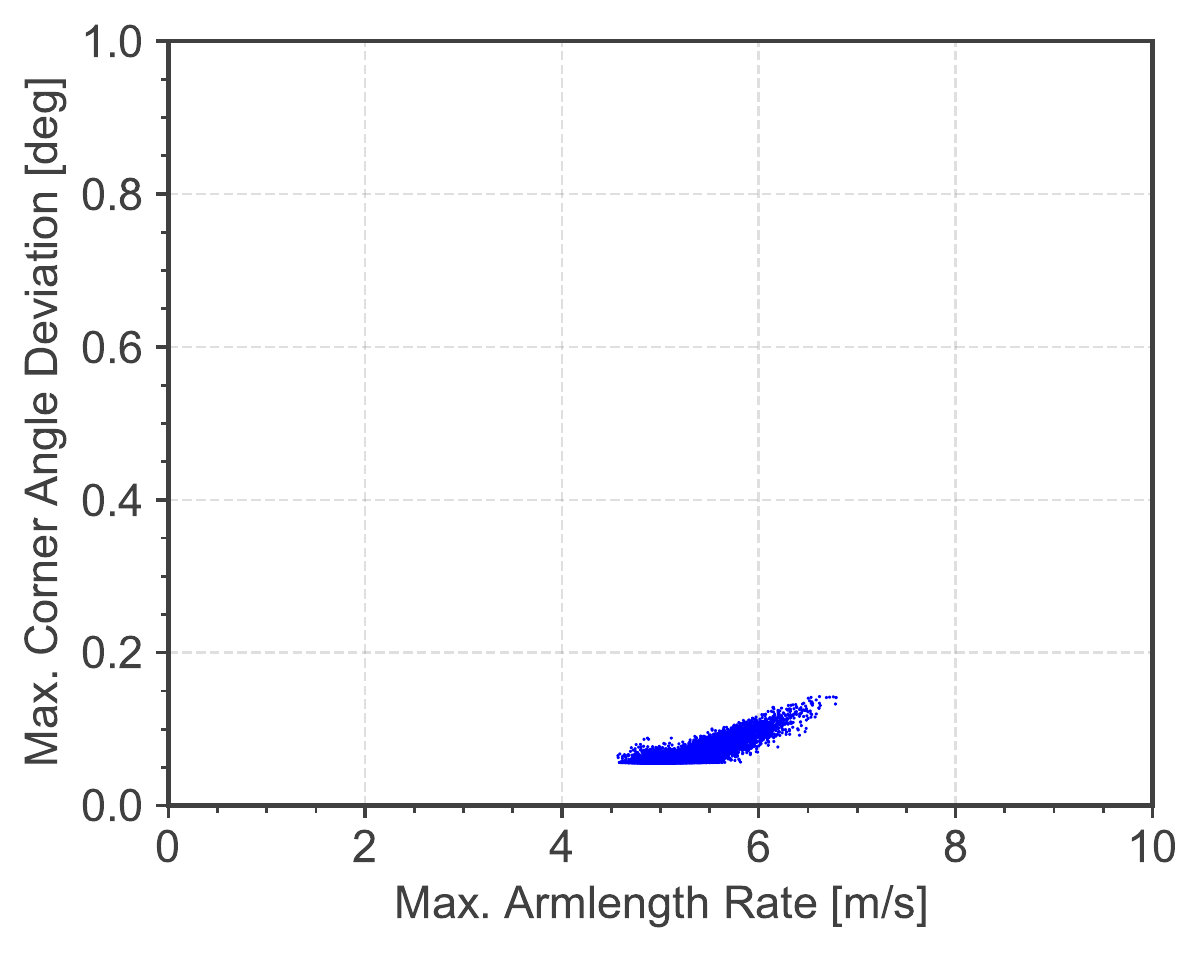}
    \includegraphics[width=7cm]{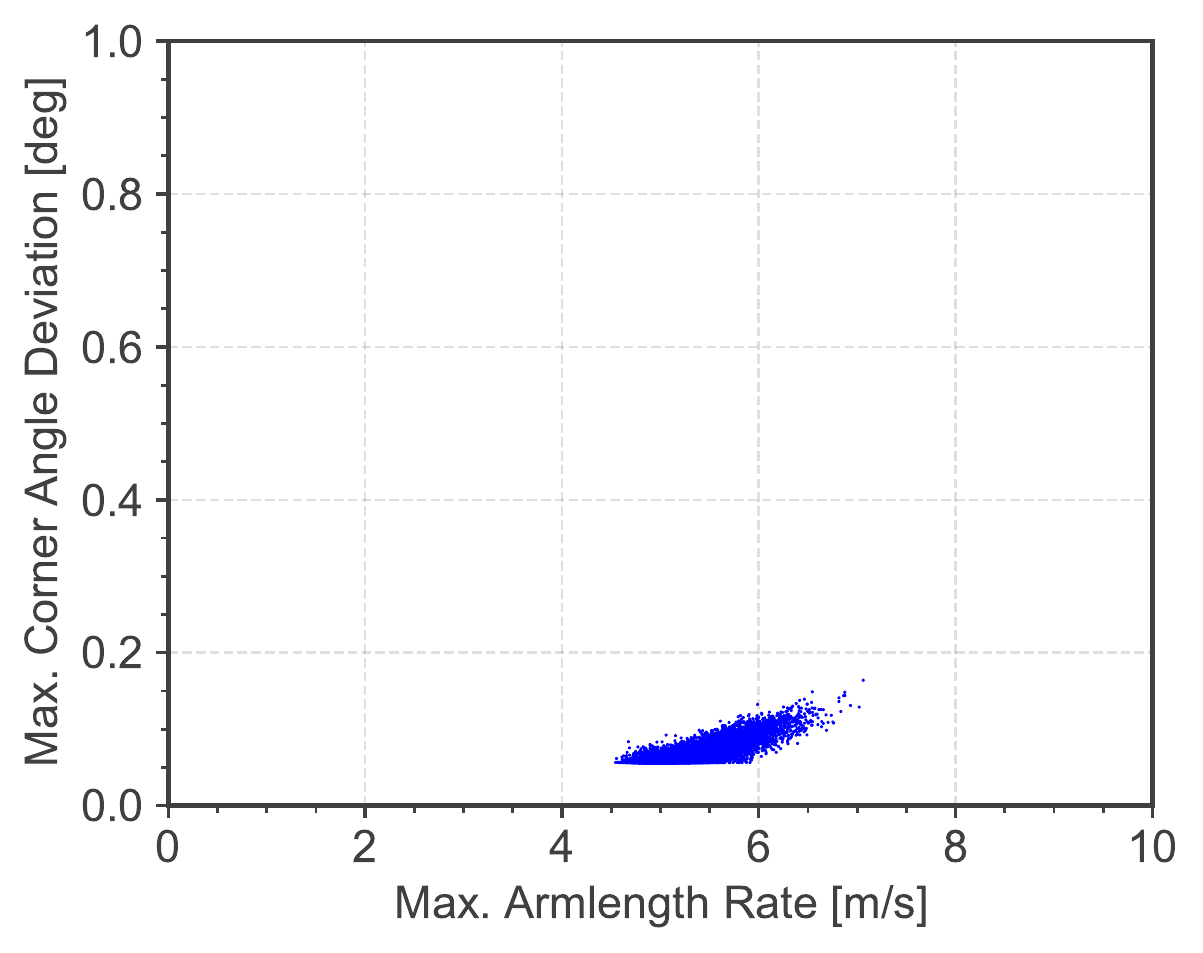}
    \begin{subfigure}{0.49\linewidth}
        \centering
        \caption{}  
    \end{subfigure}
    \begin{subfigure}{0.49\linewidth}
        \centering
        \caption{}  
    \end{subfigure}
    \caption{Sample clouds (\num{10000} samples) showing the science orbit stability due to insertion dispersion without (a) and with (b) stochastic self-gravity-induced accelerations. The shown parameter space represents the allowed region due to the assumed requirements (\SI{60+-1.0}{\degree} for the corner angles and \SI{+-10}{\meter\per\second} for the arm length rates).}
    \label{fig:insertion-mc}
\end{figure}

The science orbit stability analysis in section~\ref{sec:science-orbit} assumed a perfect insertion of the three satellites into the desired orbits. In reality, insertion disturbances of the final thruster burns are expected to disperse the initial spacecraft states. This will lead to a slight deterioration of the achievable corner angles and arm length rates. To quantify the insertion dispersion, a full simulation of the ground-based Range/Doppler tracking and guidance strategy during the transfer would be required; however, this is beyond the scope of this paper. Instead, we take some conservative assumptions on the initial state dispersion of the satellites based on experience. We perform a Monte-Carlo analysis propagating \num{10000} samples, each with slightly different initial conditions within normal distributions detailed in \cref{tab:insertion-dispersion}. These distributions are realistically achievable with standard Range/Doppler observations and a sequence of gradually decreasing clean-up maneuvers after the nominal science orbit insertion. For each sample, the maximum deviation of the corner angles and arm length rates is recorded. As shown in \cref{fig:insertion-mc} (a), the resulting dispersion on these parameters is rather limited despite the conservative assumptions, owing to the quiet environment in the vicinity of the libration points. Note that for LISA the resulting disturbance is more significant~\cite{martens2021trajectory}.

\begin{table}
    \caption{Assumed science orbit insertion uncertainties and self-gravity-induced accelerations.}
    \label{tab:insertion-dispersion}
    \begin{indented}
    \smallskip
    \scriptsize
    \item[]\begin{tabularx}{\linewidth}{@{}XllX}
    \toprule
    Parameter & Distribution & Uncertainty \\
    \midrule
    Position dispersion, $1-\sigma$\\ (along-track, cross-track, radial) & Normal & \SI{50}{km}, \SI{200}{km}, \SI{50}{km} \\
    Velocity dispersion, $1-\sigma$\\ (along-track, cross-track, radial) & Normal & \SI{10}{cm/s}, \SI{20}{cm/s}, \SI{10}{cm/s}\\
    Self-gravity-induced accelerations\\ (constant in spacecraft-fixed frame) & Uniform & \SI{2}{\nano\meter\per\second\squared} width, spherical\\
    \bottomrule
    \end{tabularx}
    \end{indented}
\end{table}

There exists a further effect, specific to drag-free spacecraft, that can destabilize the constellation due to its stochastic nature: the so-called self-gravity-induced accelerations. This is an effective acceleration component acting on the spacecraft resulting from the DFACS following the motion of the two internal test masses, if these are not located precisely at their nominal positions. To understand the effect of self-gravity-induced accelerations, first consider a single test mass that is offset from its nominal position, where the gravity acceleration induced by the spacecraft mass vanishes. Due to the gravitational force exerted by the spacecraft on the test mass, the latter will be accelerated towards its nominal location. This motion will be detected by the test-mass interferometer and the DFACS will command the propulsion system to accelerate the spacecraft into the opposite direction, such that the test mass effectively stays at rest relative to the spacecraft. This will also happen if the test mass remains at its nominal position, but the spacecraft mass distribution does not exactly cancel at this location.

This acceleration produced by the DFACS has an impact on the spacecraft trajectory. Since LISAmax is assumed to have two test masses (as LISA), the situation is more complicated in reality. For more details, refer to~\cite{martens2021trajectory}. For the purpose of the current analysis, the relevant result is that it is only possible to predict the magnitude and direction of a part of the self-gravity-induced accelerations. This known part can be taken into account in the optimization of the nominal orbit achieving similar stability. However, there is a remaining stochastic component that will act on the spacecraft and potentially destabilize the constellation. To assess its effect in the Monte-Carlo analysis, a similar assumption as for LISA has been taken on the magnitude of the acceleration (see \cref{tab:insertion-dispersion}): a uniform distribution of the three acceleration vector components defined in the spacecraft-fixed frame. The result is shown in \cref{fig:insertion-mc} (b). Again, the overall effect on the constellation stability is significantly less than for LISA. A comparison with the results obtained for LISA in reference~\cite{martens2021trajectory} is shown in \cref{tab:dispersion-statistics}. 

To summarize, neither the expected size of spacecraft insertion maneuver dispersions nor the expected levels of stochastic self-gravity-induced accelerations are sufficient to deteriorate significantly the stability of the constellation. The assumed requirements of \SI{+-1.0}{\degree} for the corner angles and \SI{+-10}{\meter\per\second} for the arm length rates are still fulfilled with significant margins when taking into account these two disturbances for the science orbit.

\begin{table}
    \caption{Results for the 99\% confidence intervals of orbit stability metrics under dispersions for LISA~\cite{martens2021trajectory} and LISAmax. Both cases include self-gravity-induced accelerations.}
    \label{tab:dispersion-statistics}
    \begin{indented}
    \smallskip
    \scriptsize
    \item[]\begin{tabularx}{\linewidth}{@{}XllX}
    \toprule
    Parameter & LISA & LISAmax \\
    \midrule
    Corner angles, 99\% C.I. & [\SI{58.507}{\degree}, \SI{61.527}{\degree}] & [\SI{59.886}{\degree}, \SI{60.101}{\degree}] \\
    Arm length rates, 99\% C.I. & [-\SI{11.04}{\meter\per\second}, \SI{10.16}{\meter\per\second}] & [-\SI{6.27}{\meter\per\second}, \SI{6.25}{\meter\per\second}]\\
    \bottomrule
    \end{tabularx}
    \end{indented}
\end{table}

\section{Sensitivity curve}
\label{sec:sensitivity}

The assumptions that we make to compute the sensitivity curve of LISAmax are summarized in \cref{tab:instrument}. A laser power of only \SI{1}{\watt} is certainly conservative; a more powerful laser could further improve the sensitivity. The driving assumption for the instrument and spacecraft design is the telescope diameter. To fully compensate for the decreased laser power at reception, one would have to scale up the telescope diameter by a factor of 10 only, compared to LISA. This is because the received power scales with the telescope diameter to the fourth power, while the travel distance only enters inversely quadratically~\cite[eq.~12]{larson2000sensitivity}:
\begin{equation}
    P_r = P_t \left[\frac{\epsilon \pi^2 \nu_0^2 D^2}{c^2} \right] \left[\frac{1}{4 \pi r^2} \right] \left[\frac{\epsilon \pi D^2}{4} \right].
\end{equation}
$P_t$ and $P_r$ are the transmitted and received power, respectively. The quantity in the first square brackets is the directional gain of the transmitter optics
with diameter $D$ and efficiency $\epsilon$ for light of frequency $\nu_0$. The quantity in the second square brackets is the space loss at a distance $r$ (arm length), and the quantity in the last square brackets is the effective cross section of the receiving optics.
Because the increased arm lengths yield an additional quadratic gain in sensitivity to gravitational-wave strain, we have increased the telescope only by about a factor of three. Nevertheless, this choice is a driver for the spacecraft design and must be carefully traded against the requirements of the science case if a LISAmax-like detector is envisioned. An increased laser power could also partly compensate for a smaller telescope diameter choice.

\begin{table}
    \caption{Instrument assumptions for LISAmax.}
    \label{tab:instrument}
    \centering
    \scriptsize
    \begin{tabularx}{\linewidth}{@{}XllX@{}}
    \toprule
    Parameter & LISA & LISAmax & Note \\
    \midrule
    Arm length [\si{\kilo\meter}] & 2.5 million & 259 million & From orbit selection. \\
    Laser wavelength [\si{\nano\meter}] & \num{1064} & \num{1064} & \\
    Laser power [\si{\watt}] & \num{1.0} & \num{1.0} & Could be increased to \SI{10}{\watt} for better sensitivity. \\
    Telescope diameter [\si{\meter}] & \num{0.3} & \num{1.0} & Scaled up to partly compensate for lower photon density. \\
    Telescope efficiency & \num{0.3}  & \num{0.3} & \\
    Test-mass noise [\si{\meter\per\second\squared\hertz\tothe{-0.5}}] & $\num{3e-15} \sqrt{1 + \left(\frac{\SI{0.4}{\milli\hertz}}{f}\right)^2}$ & $\num{3e-15} \sqrt{1 + \left(\frac{\SI{0.4}{\milli\hertz}}{f}\right)^2}$ & LISA Pathfinder achieved \SI{1.74e-15}{\meter\per\second\hertz\tothe{-0.5}} at \SI{1}{\milli\hertz}~\cite{lpf2016}. \\
    \bottomrule
    \end{tabularx}
\end{table}

\begin{figure}[h!]
    \centering
    \includegraphics[width=\linewidth]{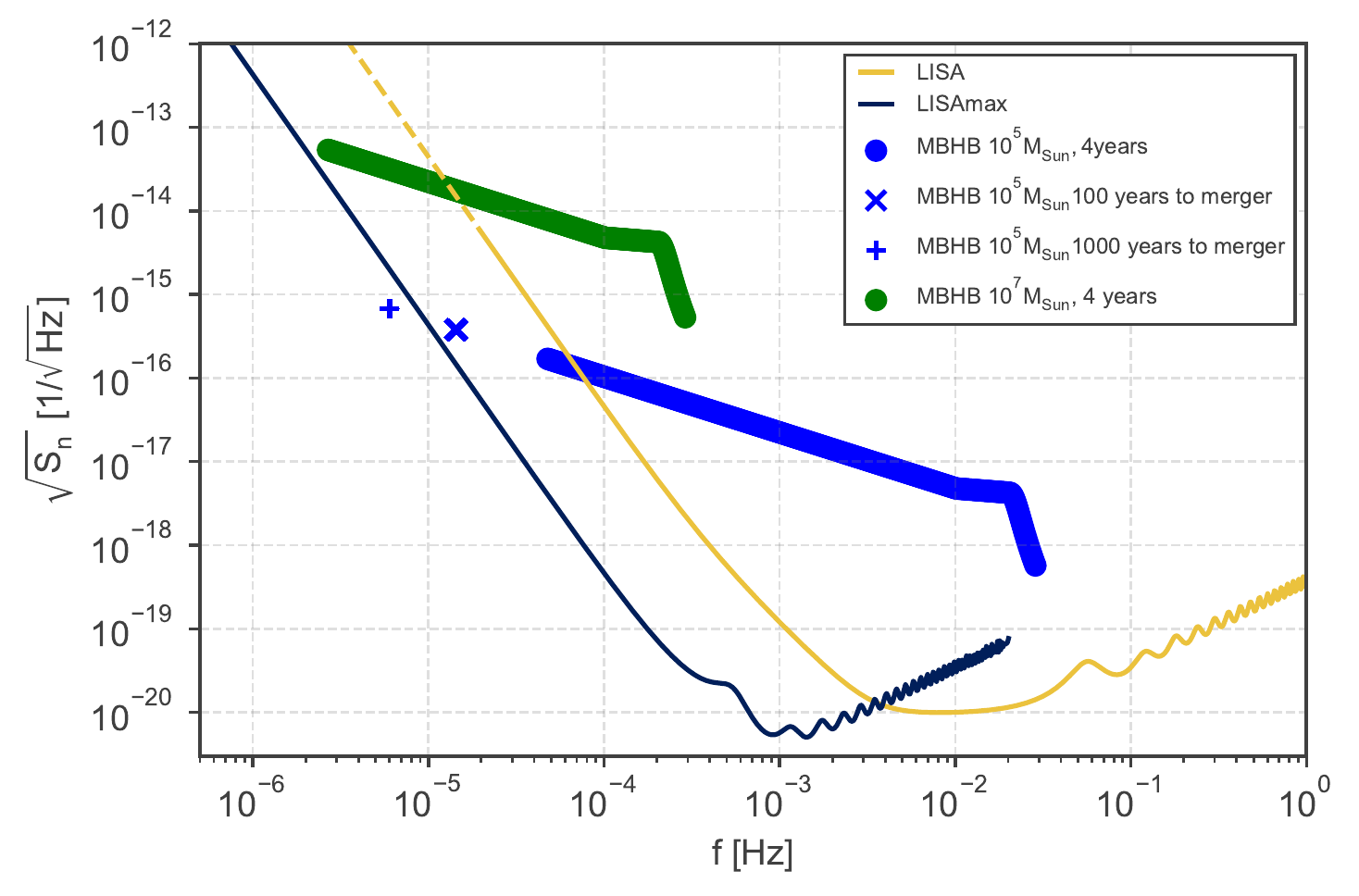}
    \caption{Sensitivity curves of LISA and LISAmax. Note that for the sake of simplicity and comparability, the same simple noise model has been used for both LISA and LISAmax~\cite{larson2000sensitivity}. As a consequence, the LISA sensitivity curve shown here is not identical to the official one. The dashed line indicates the range below the LISA curve validity. Traces of massive black hole (MBHB) inspirals and mergers at redshift $z=3$ are also shown.}
    \label{fig:sensitivity}
\end{figure}

For the sake of simplicity, the $1/f$ component of the test-mass acceleration noise has been extrapolated down to frequencies of \SI{1e-6}{\hertz}, even though the LISA Pathfinder results~\cite{lpf2016} only provide data down to \SI{2e-5}{\hertz}. It is likely that additional effects, like temperature fluctuations, alter the response at lower frequencies. Applying the same simple model~\cite{larson2000sensitivity} for both LISA and LISAmax allows for a indicative comparison. \Cref{fig:sensitivity} shows the sky-averaged sensitivity curves (Michelson combination) in amplitude spectral density for both LISA and LISAmax.

The overall downward shift of the LISAmax sensitivity curve by two orders of magnitude at low frequencies is a direct consequence of the increased constellation arm lengths: a given gravitational wave causes a stronger signal in a larger laser arm. However, the increased arm lengths also shift the right slope towards lower frequencies as the observatory becomes less sensitive to smaller wavelengths (arm length penalty). The highest sensitivity point is determined by the laser shot noise, which is inversely proportional to the received laser power. With the current assumptions, this point is slightly lower for LISAmax than for LISA (around \SI{E-3}{\hertz}, compared to LISA's \SI{E-2}{\hertz}). It can be further reduced by either increasing the telescope diameter ($S_n \propto D^{-4}$) or by increasing the transmitted laser power ($S_n \propto P_t^{-1}$). The slope on the left is determined by the test-mass acceleration noise. Note that no Galactic confusion noise is taken into account here. It degrade the sensitivity of both detectors in the area around and below \SI{1}{\milli\hertz}. A recent study~\cite{Wang:2023jct} showed that a large part of the Galactic double white dwarf population could not be resolved by a \si{\micro\hertz} detector, and would produce a confusion noise above the instrumental noise, potentially down to \SI{1e-5}{\hertz}.

For convenience, we also list the analytical formulas for the included noise power spectral densities in the LISAmax sensitivity curve consisting of shot noise and test mass acceleration noise:
\begin{subequations}
\begin{align}
    S_{n,\text{shot}} &= 2 \frac{ 4 c^4 h}{\pi^4 \epsilon^2 D^4 P_t \nu_0^3}, \\
    P_\text{acc} &= \left(\SI{3e-15}{\meter\per\second\squared}\right)^2 \left(1 + \left(\frac{\SI{0.4}{\milli\hertz}}{f}\right)^2\right) \si{\per\hertz}, \\
    S_n &= S_{n,\text{shot}} + \frac{4}{(2 \pi f)^4 L^2} P_\text{acc},
\end{align}
\end{subequations}
where $L$ is the arm length of the detector and the other variables have been defined previously.

A note regarding directional sensitivity is in order here. LISA’s constellation axis precesses around the ecliptic pole with an opening half angle of \SI{60}{\degree} once per year, while the constellation is also rotating around its own axis. This leads to an amplitude modulation of a long-lived gravitational-wave signal from a fixed location in the sky over the year (due to LISA's non-spherical antenna pattern). In the case of a LISAmax-like detector, the constellation normal does not precess because the constellation is bound to the ecliptic plane. Therefore, localizing a gravitational-wave source in the sky cannot rely solely on this amplitude modulation.

There are two main ways to mitigate this:
\begin{enumerate}
    \item Incline each satellite orbit with respect to the ecliptic, but offset the three nodes by \SI{120}{\degree}~\cite{Wang_2015,sesana2021unveiling}. This will cause a precession of the constellation normal around the ecliptic pole in way similar to LISA. As a consequence, equivalent sky localization methods could be employed.
    \item Add a second constellation with a significant relative inclination, as proposed in reference~\cite{sesana2021unveiling}. Both constellations will simultaneously measure a given gravitational-wave signal with different amplitude modulations. This would allow narrowing down the sky location, possibly up to an ambiguity.
\end{enumerate}
Both options will be discussed in more detail in \cref{sec:options}. However, they would both imply a significant leap in mission complexity and cost compared to the baseline concept presented here.

In addition to the precession of the constellation plane, LISA's center moves around the Sun, causing a seasonal Doppler modulation of its response. This effect is missing in LISAmax. Note that, however, because the LISAmax constellation rotates around the Sun once per year, there is still a seasonal variation of the directional sensitivity, which can be exploited to gain some information on the source location. Moreover, different time-delay interferometry (TDI) combinations can be constructed from the raw Doppler readouts; each of these combinations has a different antenna pattern~\cite{tinto1999cancellation,Armstrong_1999,tinto2014time}. This can also help to obtain information about the direction of the gravitational-wave source. Therefore, we do not expect that LISAmax looses all resolution power. A more in-depth study of the localization capabilities of LISAmax-like missions is being prepared.

To illustrate one science case of LISAmax, \cref{fig:sensitivity} also shows the traces of massive black hole inspirals and mergers in the sensitivity plot~\cite{robson2019construction}. While the merger itself for both \num{E5} and \num{E7} solar mass black-hole binaries is fully within the LISA band, LISAmax can observe such objects dozens of years before the actual merger.

\section{Interplanetary transfer}
\label{sec:transfer}

A main driver for the system design is the fuel needed for the transfer. This can be computed from the transfer $\Delta v$. The following assumptions have been made for the interplanetary transfer optimization:
\begin{itemize}
    \item Each spacecraft is launched on a dedicated European launcher from Kourou with an escape velocity of \SI{1}{\kilo\meter\per\second}. This assumption is a major cost driver and depends mainly on the choice of the telescope size, and thus the spacecraft mass. With a smaller telescope than assumed here (cf.,~\cref{tab:instrument}), a shared launch of all three satellites, similar to LISA, would be conceivable.
    \item At \SI{1}{\kilo\meter\per\second} escape velocity, the launcher is assumed to deliver 10 tons of payload. This is in line with the expected developments of the large European launch vehicle, Ariane 6.4~\cite{lagier2021ariane}. It would also be well within the capabilities of Falcon Heavy or Vulcan Centaur 6. 10 tons should allow for accommodation of a telescope of the size assumed here.
    \item The transfer design assumes chemical propulsion with high thrust, which allows the maneuvers to be approximated by instantaneous velocity increments.
\end{itemize}

\begin{figure}[h!]
    \centering
    \includegraphics[width=\linewidth]{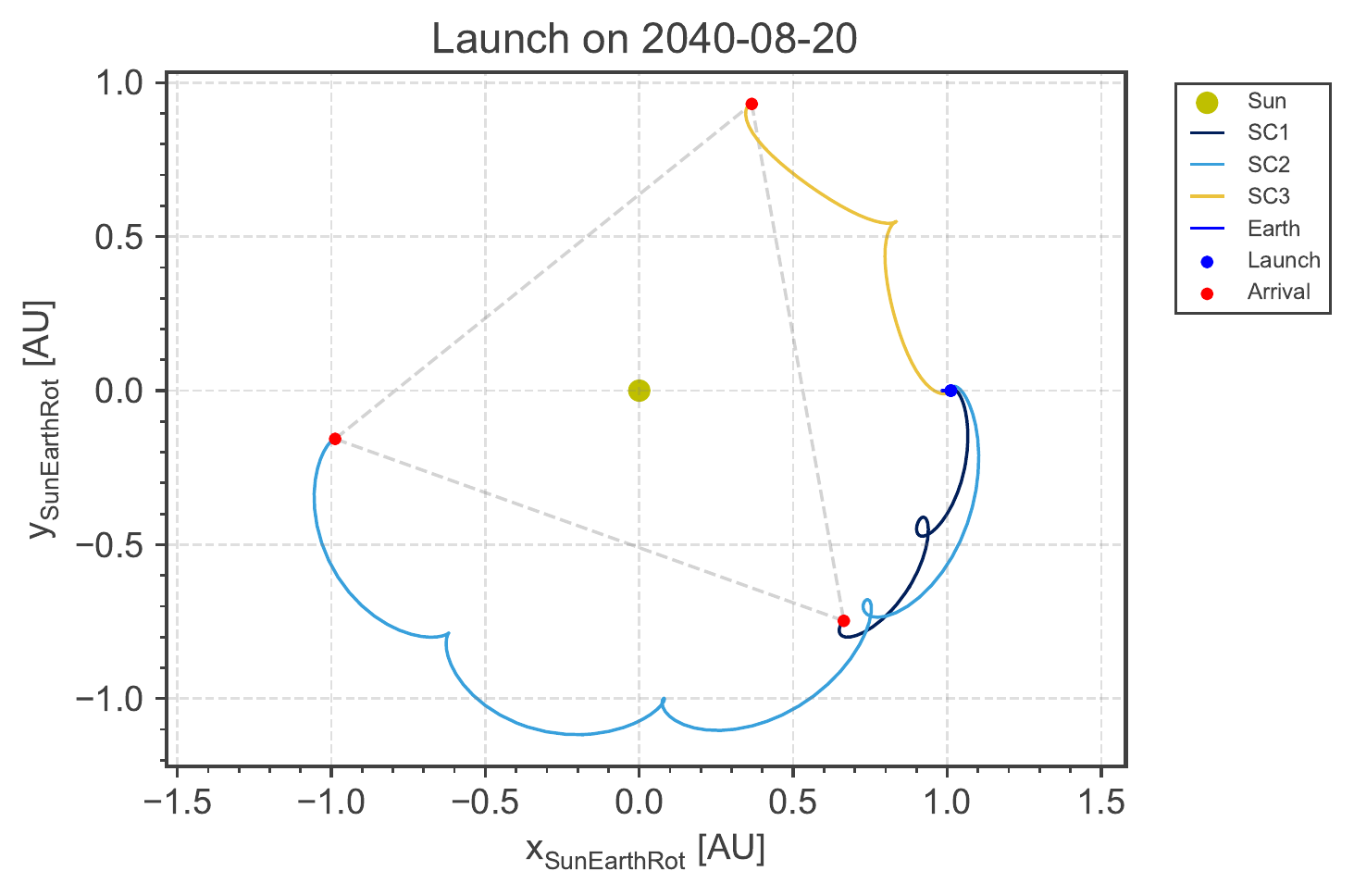}
    \caption{Transfer trajectories using dedicated launches in the Sun-Earth rotating frame.}
    \label{fig:aug_traj2d_rot}
\end{figure}

To transfer a spacecraft from Earth to the desired location, close to the triangular libration points, a simple phasing strategy can be employed: launch injects the spacecraft into a heliocentric transfer orbit. To reach the near-L4 region, the transfer is a heading orbit with a semi-major axis below \SI{1}{\astronomicalunit}. To reach the near-L3 and near-L5 regions, the transfer is a trailing orbit with a semi-major axis above \SI{1}{\astronomicalunit}. When the desired location is reached, a maneuver is applied to insert into the circular target orbit and stop the drift relative to Earth. The required transfer $\Delta v$ is a function of the transfer duration, which can be a multiple of 1 year, minus (for L4) or plus (for L3 and L5) a fixed increment. The longer the transfer and the more heliocentric revolutions are completed, the lower the $\Delta v$. From the cost perspective, it is advantageous to have three spacecraft of identical design, which means that all three have the same $\Delta v$ capacity. This can be achieved assuming two revolutions around the Sun for spacecraft 1 and 3 (going to L4 and L5) and four revolutions for spacecraft 2. The fully optimized transfer trajectories for a launch in August 2040 are shown in \cref{fig:aug_traj2d_rot}.

\begin{figure}[h!]
    \centering
    \includegraphics[width=16cm]{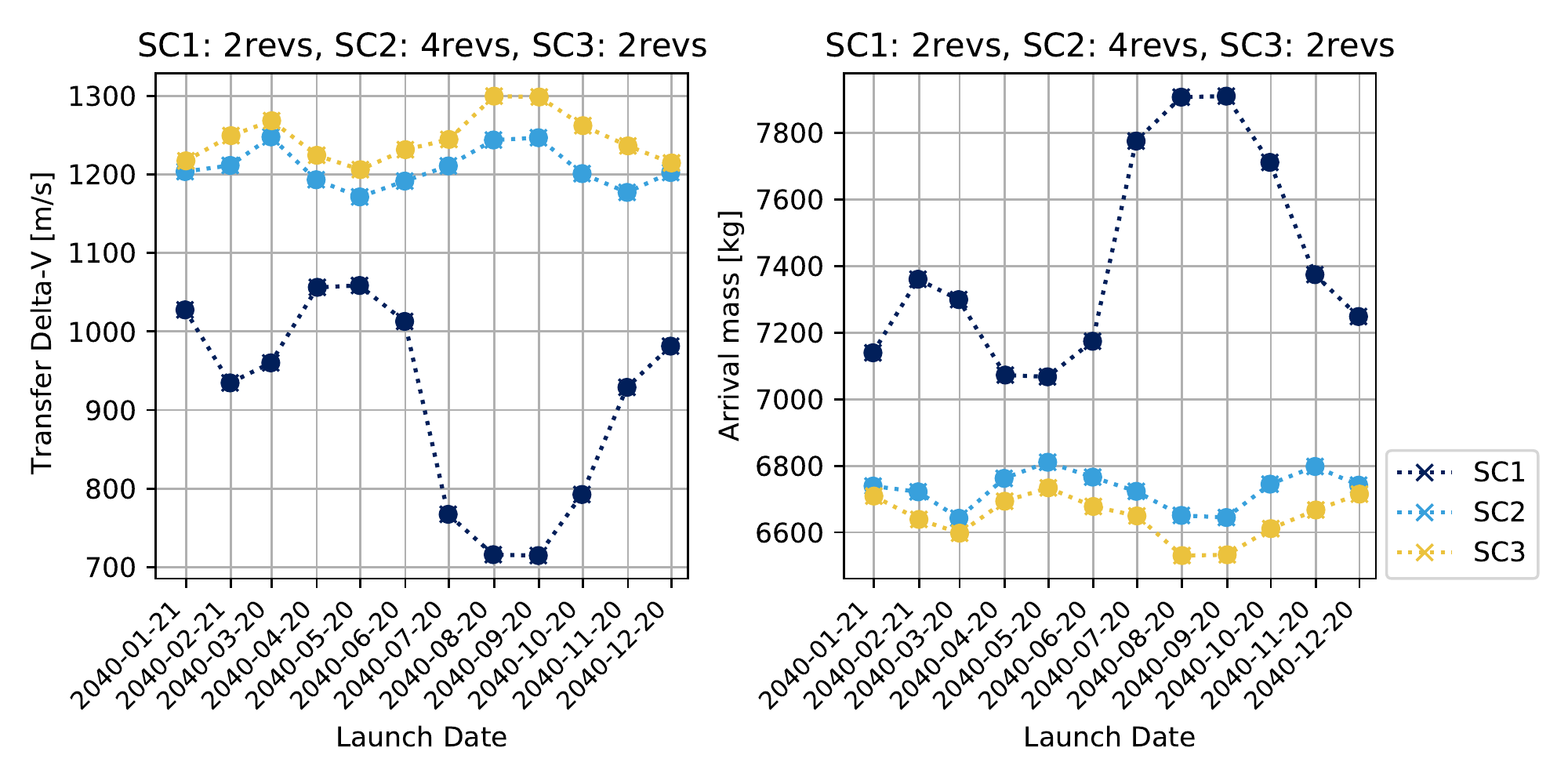}
    \begin{subfigure}{0.49\linewidth}
        \centering
        \caption{}  
    \end{subfigure}
    \begin{subfigure}{0.49\linewidth}
        \centering
        \caption{}  
    \end{subfigure}
    \caption{Full year transfer $\Delta v$ (a) and spacecraft mass in science orbit (b).}
    \label{fig:dv_and_mass}
\end{figure}

The computation of these transfers was set up as a multiple shooting problem with a Lambert arc as an initial guess. The spacecraft mass placed into final orbit was used as the objective function of the optimization problem. Due to the eccentricity of the Earth orbit, the transfer $\Delta v$, and thus the arrival mass, depends on the launch season. The results including the seasonal variations over the full year of 2040 are shown in \cref{fig:dv_and_mass}. The results, however, are applicable to any launch year.

Spacecraft 1 (going to L5) has the lowest $\Delta v$. This is because it has the shortest travel distance and the target orbit is Earth-trailing, which has a slight advantage $\Delta v$-wise. Spacecraft 2 (going to L3) has almost as large a $\Delta v$ as spacecraft 3 (going to L4) in spite of a much longer transfer duration. This is because spacecraft 2 has the longest travel distance. Assuming that launch can take place on any day of the year, the minimum mass inserted into any of the target locations still amounts to \SI{6500}{\kilogram}. This is about three times larger than that of LISA. It should be sufficient to accommodate a larger telescope, the larger high-gain antenna and include the necessary support structure.

\begin{figure}[h!]
    \centering
    \includegraphics[width=\linewidth]{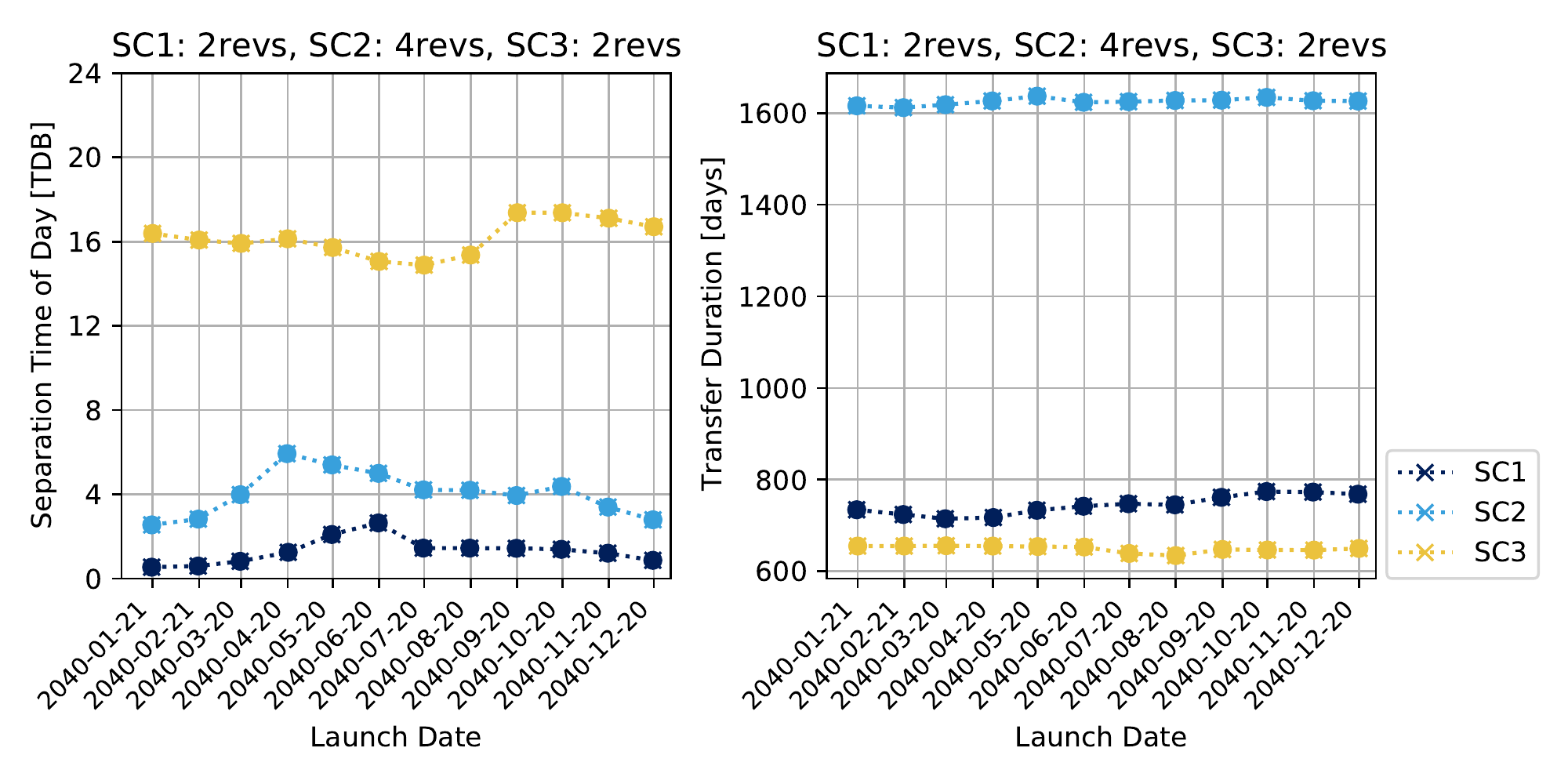}
    \begin{subfigure}{0.49\linewidth}
        \centering
        \caption{}  
    \end{subfigure}
    \begin{subfigure}{0.49\linewidth}
        \centering
        \caption{}  
    \end{subfigure}
    \caption{Full year launcher separation TDB hour (a) and transfer duration (b).}
    \label{fig:separation_and_duration}
\end{figure}

The TDB hour of the launcher separation as well as the transfer duration for the full launch period are shown in \cref{fig:separation_and_duration}. The different escape directions targeted by the three spacecraft result in different launch times (and hence launcher separation times) because the launcher trajectory is fixed in the Earth-fixed frame.  Spacecraft 2 has a transfer duration of around 4.5 years. If consecutive launches are envisioned, this one should be launched first to minimize the waiting time of the others before full science operations can commence. This may also help to spread the construction effort and cost over a longer time period. 

\section{Mission options}
\label{sec:options}

One of the main challenges of an \si{\astronomicalunit}-sized detector, such as the one studied here, is to preserve a good gravitational-wave source localization capability. In the LISA mission, for long-lived sources, this is achieved by tracking the amplitude of the response to the gravitational-wave signal of a given source over the course of the year. The precession of the constellation plane around the ecliptic north pole induces a characteristic amplitude modulation in the gravitational-wave signal that can be used to estimate the sky location. Moreover, the motion of the constellation around the Sun leads to a location-dependent Doppler shift when the detector is moving towards or away from the source. Both of these effects are missing in LISAmax. The only seasonal variation is due to the rotation of the full constellation around the Sun once per year.
In this section, several mission options that address the challenge of sky localization and the possibility of a joint launch are discussed.

\subsection{Precessing inclined orbits}
\label{sec:precessing}

\begin{figure}[h!]
    \centering
    \begin{subfigure}{0.49\linewidth}
        \raisebox{1.5cm}{\includegraphics[width=6.5cm]{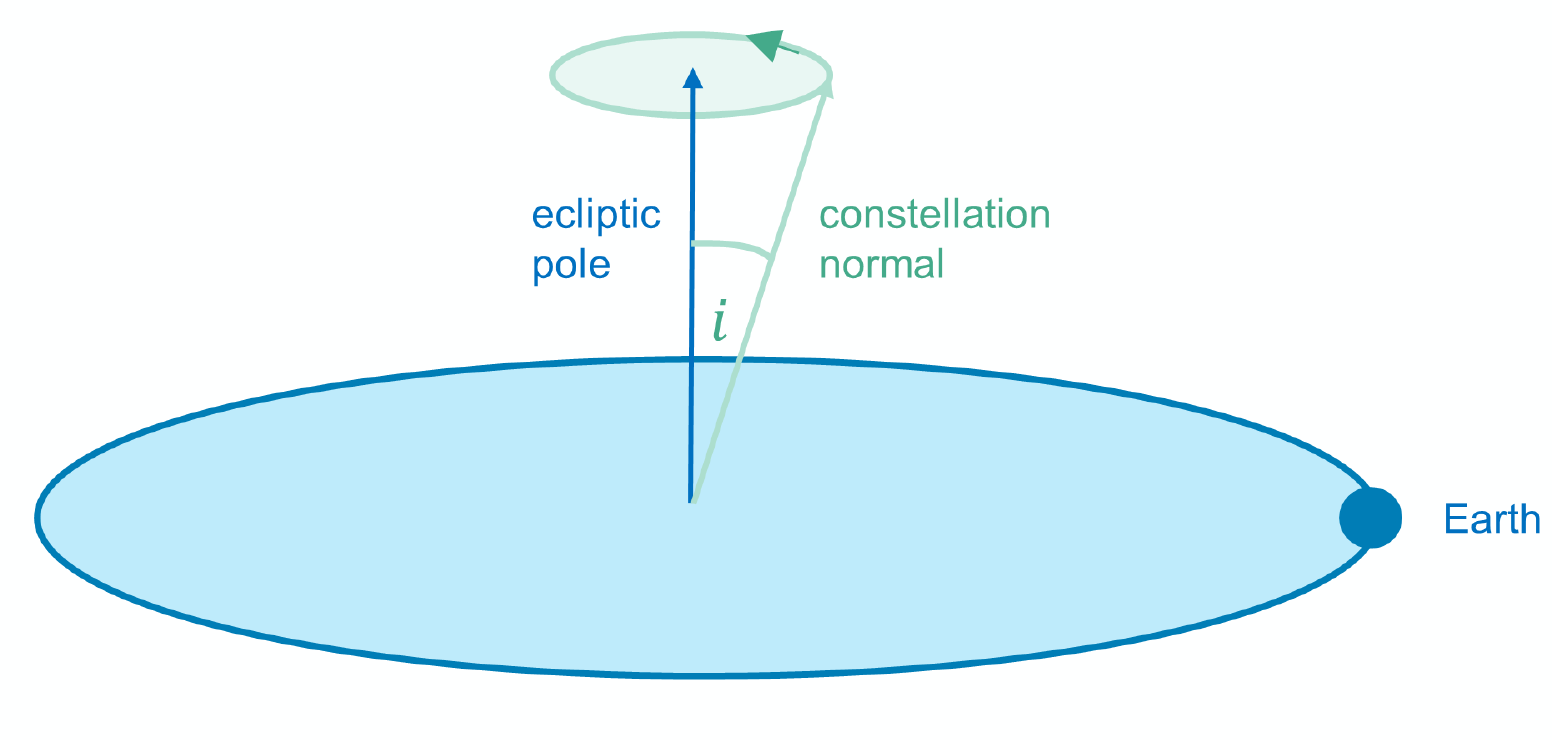}}
    \caption{}
    \end{subfigure}
    \begin{subfigure}{0.49\linewidth}
        \includegraphics[width=8cm]{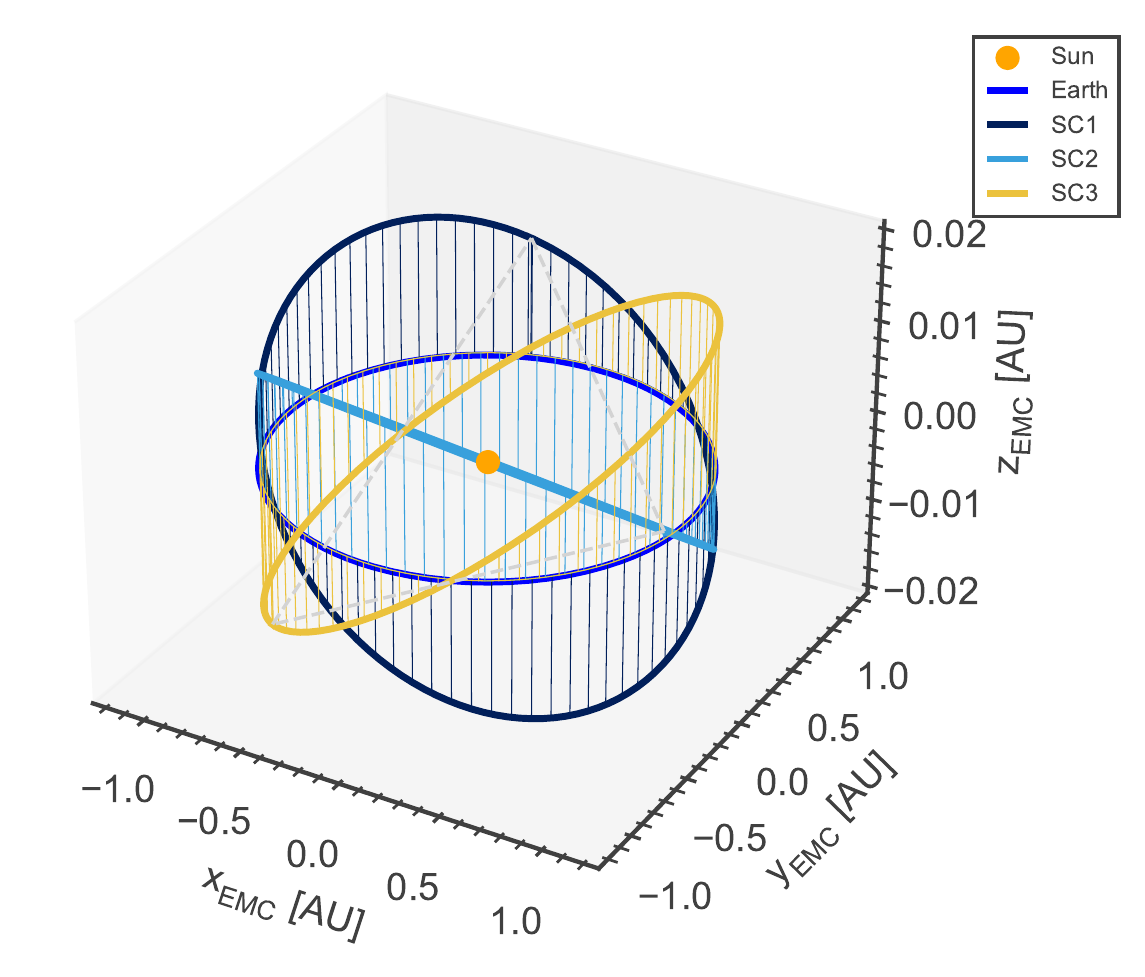}
         \caption{}
    \end{subfigure}
    \caption{Illustration of precessing inclined orbits (a). (b) shows orbits with an inclination of \SI{1}{\degree} relative to the ecliptic.}
    \label{fig:precession}
\end{figure}

A precessing motion similar to that of LISA can be introduced to LISAmax by inclining all three orbits with respect to the ecliptic and spacing the nodes of the orbital planes by \SI{120}{\degree}~\cite{Wang_2015,sesana2021unveiling}. This is illustrated in \cref{fig:precession} where the optimized orbits in the full numerical model (c.f., \cref{sec:science-orbit}) are shown. The constellation normal will precess around the ecliptic pole with a period of one year and induce a variation in the gravitational-wave response amplitude over the course of a year similar to that of LISA.

\begin{figure}[h!]
    \centering
    \begin{subfigure}{\linewidth}
        \includegraphics[width=\linewidth]{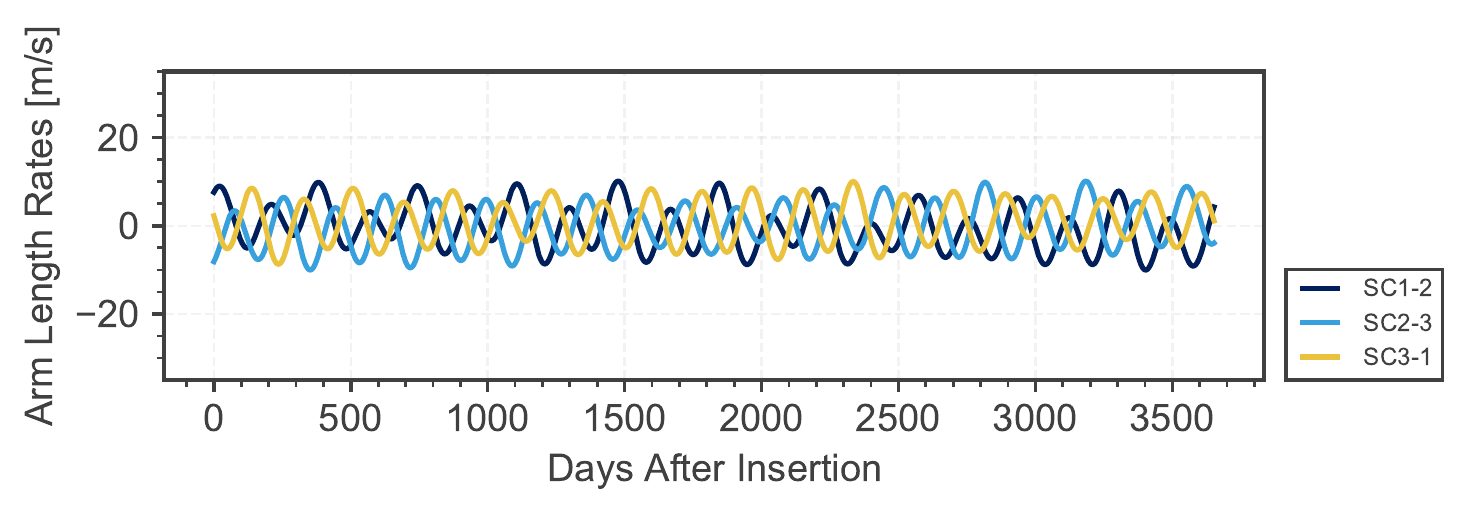}
        \caption{}
    \end{subfigure}
    \begin{subfigure}{\linewidth}
        \includegraphics[width=\linewidth]{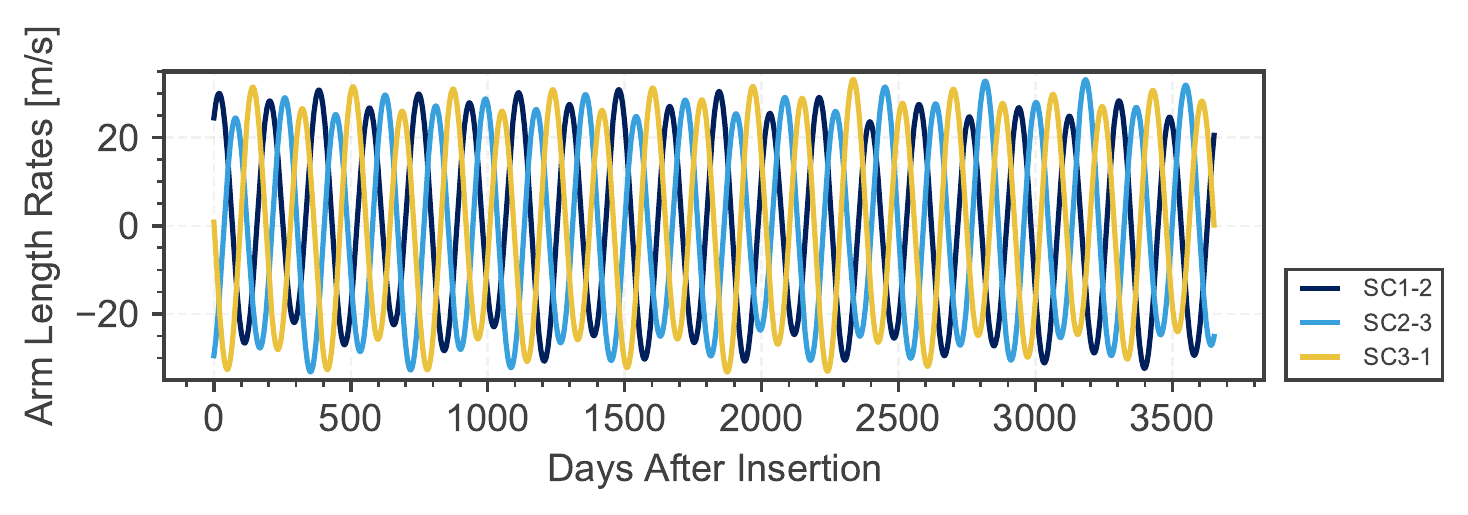}
        \caption{}
    \end{subfigure}
    \caption{Arm length rate for precessing inclined orbits with \SI{1}{\degree} (a) and \SI{2}{\degree} (b) ecliptic inclination. The arm length rates increase strongly, the higher the inclination.}
    \label{fig:armlength_rate_precessing}
\end{figure}

One problem with this kind of precessing orbits is that the arm length rate between the satellites increases with the chosen inclination. This is illustrated in \cref{fig:armlength_rate_precessing}. An inclination of \SI{2}{\degree} already leads to an arm length rate variation of $\pm$\SI{33}{\meter\per\second} (to be compared with $\pm$\SI{5}{\meter\per\second} without inclination). Attempts have been made to reduce the breathing of the triangle by assuming a non-zero eccentricity, but were unsuccessful as the optimizer converges to circular orbits in all cases. Future work will focus on obtaining the optimal eccentricities and arguments of perihelion in the Keplerian model analytically to confirm that circular orbits are indeed optimal. In the full numerical model, this appears to be the case.

Therefore, to use precessing orbits with inclinations higher than \SI{1}{\degree}, one must cope with the ensuing relative velocities of significantly more than \SI{10}{\meter\per\second}. This can only be done by increasing the bandwidth of the phasemeter, which is currently a technological limitation (large velocities imply large Doppler shifts and therefore a large variation of spacecraft beatnote frequencies).

\subsection{Planetary flybys to reach high inclinations}
\label{sec:flybys}

A mission concept with a combination of two triangular constellations has been proposed in~\cite{sesana2021unveiling}, where one is inclined by \SI{90}{\degree} relative to the ecliptic while the other one remains in the ecliptic. However, reaching such high inclinations requires a prohibitive transfer $\Delta v$: to transfer from an initial Earth-like orbit to one inclined by \SI{90}{\degree}, an impulsive $\Delta v$ of about $\sqrt{2} \cdot 30$ \si{\kilo\meter\per\second} has to be imparted. This is beyond the capability of currently known rocket engines.

\begin{figure}[h!]
    \centering
    \includegraphics[width=\linewidth]{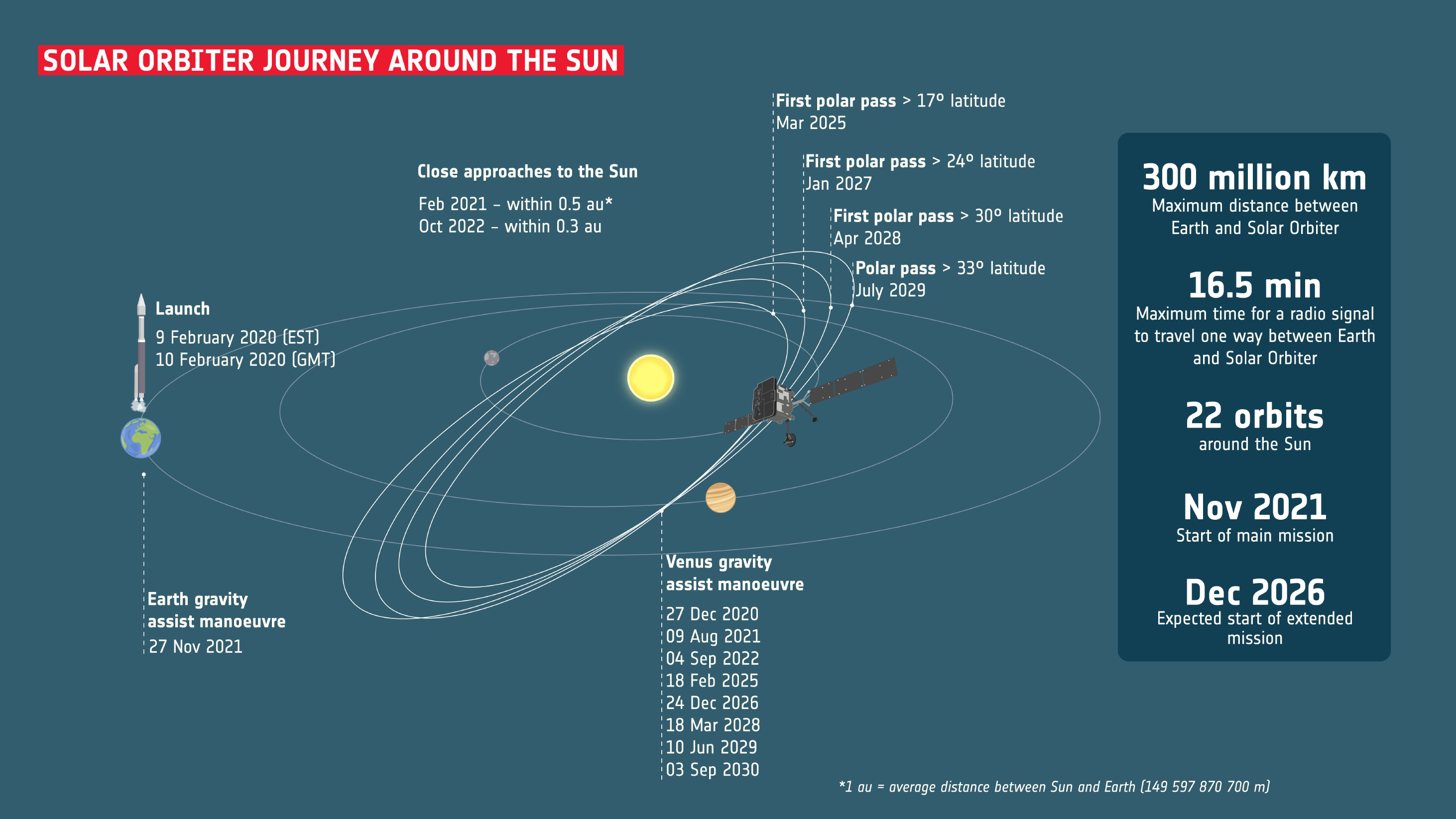}
    \caption{Solar Orbiter trajectory. High solar latitudes are reached using a sequence of Venus flybys. Image retrieved from~\cite{solo-image}.}
    \label{fig:solar_orbiter}
\end{figure}

To achieve a significant relative inclination between two constellation planes, however, planetary flybys can be exploited. For instance, the currently operational Solar Orbiter mission uses a series of resonant Venus flybys to reach a final inclination of \SI{33}{\degree} (extended mission) relative to the Sun equator in order to observe the Sun from its poles~\cite{perez2012trajectory,perez2018solar,marirrodriga2021solar} (see \cref{fig:solar_orbiter}). In principle, a similar strategy with Venus and Earth flybys could be employed to place two constellations with opposite inclination values, such that the relative inclination adds up. However, such a mission would have a vastly higher complexity and cost compared to the baseline concept presented here. Not only is the transfer for a single spacecraft significantly longer and more complex, but all six spacecraft would have to follow dedicated flyby trajectories, increasing also the operational cost. Such a concept could only be facilitated by a significant increase in spacecraft autonomy compared to what is customary at the present day.

\subsection{Constellations beyond 1~au}

A simple modification to the baseline LISAmax concept is to increase the arm length even further by placing the satellites on a circular orbit beyond \SI{1}{\astronomicalunit}, for instance at the altitude of the Mars orbit, as suggested in~\cite{sesana2021unveiling}. That would further improve the sensitivity at low frequencies, but would also come with the following downsides:
\begin{itemize}
    \item The transfer $\Delta v$ increases significantly due to the higher Earth escape velocity and arrival maneuver required to reach larger orbits.
    \item The stable geometry relative to the Sun and the Earth would be lost due to the different orbital periods of the Earth and the satellites around the Sun. This would require a regular antenna re-orientation to communicate with the Earth, which induces undesirable noise on the spacecraft. Moreover, all three spacecraft would have to be designed to cope with the maximum Earth distance. In the baseline concept, only the satellite located close to L3 needs to communicate over the large distance of almost \SI{2}{\astronomicalunit}.
    \item Due to the relative drift between the Earth and the spacecraft, regular solar conjunctions would occur, which would interrupt two-way communications with the Earth for up to several weeks.
\end{itemize}

Overall, these significant downsides must be weighed against the relatively limited gain in sensitivity and no gained advantage in source localization power. It appears that tuning other key parameters, such as the laser power, is a more achievable means of increasing the sensitivity. 

\subsection{Joint launch}

\begin{figure}[h!]
    \centering
    \includegraphics[width=0.5\linewidth]{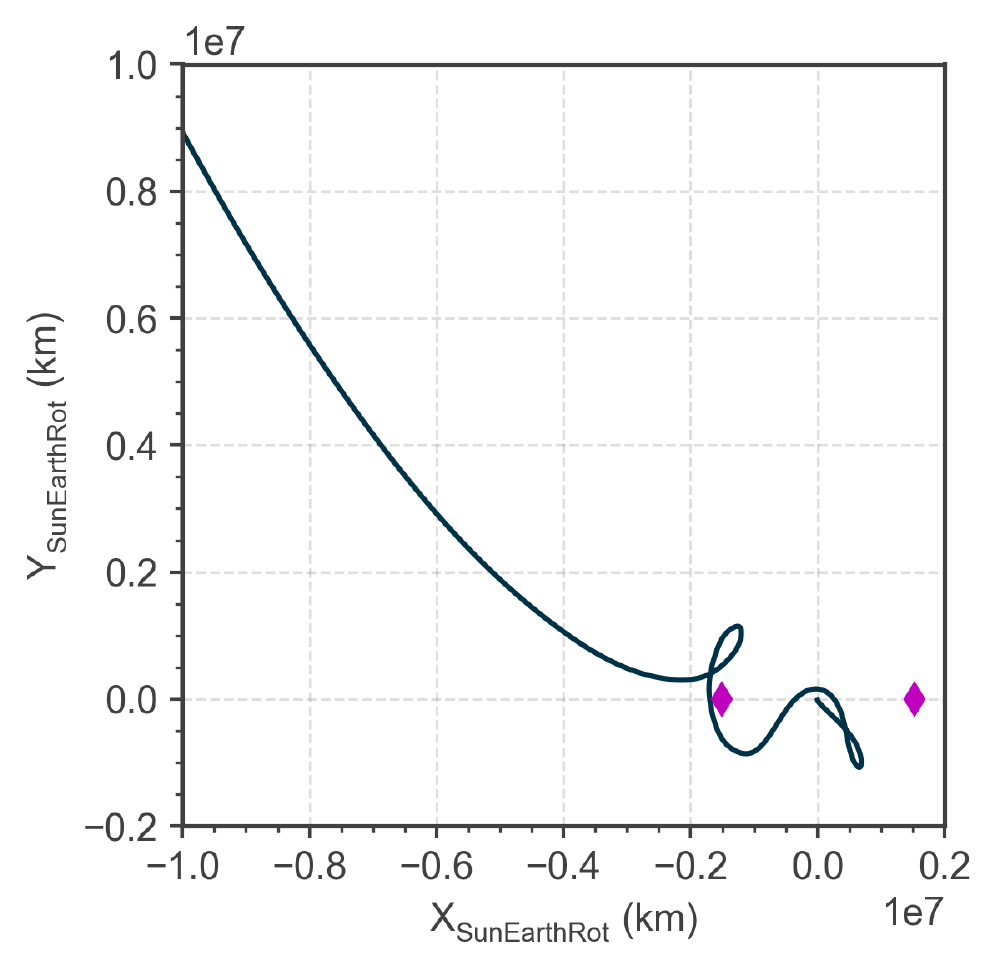}
    \caption{An example of a heteroclinic connection between L2 (right diamond) and L1 (left diamond) in the Sun-Earth rotating frame with the Earth in the origin. Using such a trajectory can facilitate an escape from Earth towards the Sun, even though the launch was away from the Sun. Image courtesy of Lorenzo Bucci.}
    \label{fig:heteroclinic}
\end{figure}

The dedicated launch scenario (a separate rocket for each satellite) considered in this paper is driven by the assumption of a large telescope mass (\SI{1}{\meter} in diameter) and the resulting large spacecraft mass. A joint launch (all three satellites on the same rocket) would likely be favorable from the mission cost point of view if a smaller telescope is used. It would, however, complexify the transfer strategy, because the Earth departure velocity asymptote can no longer be chosen separately for the three spacecraft. While spacecraft 1 and 2 require a launch away from the Sun, into an Earth-trailing orbit, spacecraft 3 requires to be launched towards the Sun into an Earth-heading orbit. In a joint launch scenario this can be achieved, for instance, by launching towards the Sun-Earth libration point L2 and exploiting the heteroclinic connections towards L1 with spacecraft 3 only (see \cref{fig:heteroclinic}). Each spacecraft would have to perform small correction maneuvers shortly after launch to acquire the respective escape trajectories. A similar strategy will be employed by the ESA Vigil mission~\cite{palomba2022vigil}, which takes the opposite route (from L1 towards L2).

An alternative joint launch strategy would employ lunar flybys to separate the Earth-leading and Earth-trailing paths of the spacecraft, similar to the NASA STEREO mission~\cite{dunham2009stereo}. This is more complex in terms of trajectory design and spacecraft navigation, but may have a higher achievable dry mass.

\section{Conclusions and outlook}
\label{sec:conclusions}

A mission concept for a space-based gravitational-wave observatory with a sky-averaged sensitivity two orders of magnitude better at low frequencies than that of LISA has been studied. A LISAmax-like detector could be a potential successor of LISA as part of ESA's Voyage 2050 planning cycle. The analysis of the operational orbit shows a better stability in terms of corner angle variations and arm length rates over 10 years than that of LISA without station keeping. Taking into account the effect of insertion dispersions and self-gravity-induced accelerations does not significantly alter that picture. With three dedicated launches using a future European launcher from Kourou, a mass into science orbit of \SI{6500}{\kilogram} per spacecraft (about three times the LISA mass) could be achieved. Such a large spacecraft mass is mainly motivated by the assumption of a larger telescope, assumed to be \SI{1}{\meter} in diameter (compare this to \SI{30}{\centi\meter} for LISA). It is not clear whether such a large telescope is actually needed for a compelling science case. A smaller telescope would significantly reduce the spacecraft mass and might permit a cheaper joint launch. It would also deteriorate the instrument sensitivity by increasing the laser shot noise, but the loss would mainly be in the region around \SI{1}{\milli\hertz}, where Galactic confusion noise is expected to be dominant in any case.

The main challenge of the presented mission concept is to retain the ability to localize gravitational-wave sources, since the constellation is constrained to the ecliptic plane. A dedicated in-depth study of the LISAmax sky localization capability is underway. Several mitigation options have been preemptively discussed in the paper. The considered mission options also show that LISAmax is the largest constellation that can be achieved without another leap in mission complexity and cost. Therefore, it is the best starting point when considering a space-based gravitational-wave detector for the \si{\micro\hertz} band.

Overall, no show stoppers are identified so far. A LISAmax-like detector would be a valuable continuation of the LISA program, by extending the sensitive frequency band below the LISA band and allowing observations of super-massive black hole inspirals many of years before the merger, and a vast population of Galactic binary systems. Studies are ongoing to characterize the full science case of a LISAmax-like \si{\micro\hertz} gravitational-wave observatory.

\ack The authors would like to thank Oliver Jennrich for valuable discussions. The image in \cref{fig:heteroclinic} was kindly provided by Lorenzo Bucci. Jean-Baptiste Bayle gratefully acknowledges support from UK Space Agency via STFC [ST/W002825/1].

\section*{References}

\bibliographystyle{iopart-num}
\bibliography{bibliography}

\end{document}